\documentclass[a4paper,11pt]{article}
\usepackage{jheppub} 
\makeatletter
\def\@fpheader{\relax}
\makeatother

\newcommand{\C}{\mathbb{C}}
\newcommand{\N}{\mathbb{N}}

\newcommand{\Z}{\mathbb{Z}}

\newcommand{\abracket}[1]{\left\langle#1\right\rangle}
\newcommand{\bbracket}[1]{\left[#1\right]}

\newcommand{\bracket}[1]{\left(#1\right)}

\newcommand{\bl}{\textbf}

\newcommand{\mc}{\mathcal}

\newcommand{\pa}{\partial}

\newcommand{\into}{\hookrightarrow}

\newcommand{\iso}{\cong}

\newcommand{\0}{\mathbf{0}}

\newcommand{\xx}{\bl{x}}

\renewcommand{\Im}{\operatorname{Im}}

\DeclareMathOperator{\Res}{Res}
\DeclareMathOperator{\Jac}{Jac}
\DeclareMathOperator{\NN}{\mathbb{N}}

\def\beq#1\eeq{\begin{align}#1\end{align}}

\title{Seiberg-Witten differential via primitive forms}

\author[d]{Si Li}
\author[b,c]{Dan Xie}
\author[a,b,c]{Shing-Tung Yau}

\affiliation[a]{Department of Mathematics, Harvard University, Cambridge, MA 02138, USA}
\affiliation[b]{Center of Mathematical Sciences and Applications, Harvard University, Cambridge, 02138, USA}
\affiliation[c]{Jefferson Physical Laboratory, Harvard University, Cambridge, MA 02138, USA}  
\affiliation[d]{Department of Mathematics and YMSC, Tsinghua University, Beijing, 100084, China}

\abstract{Three-fold quasi-homogeneous isolated rational singularity is argued to define a four dimensional $\mathcal{N}=2$ SCFT. The Seiberg-Witten geometry 
is built on the mini-versal deformation of the singularity. We argue in this paper that the corresponding Seiberg-Witten differential is given by the Gelfand-Leray form of K. Saito's primitive form. Our result 
also extends the Seiberg-Witten solution to include irrelevant deformations.}

\begin{document} 
\maketitle
\flushbottom

\section{Introduction}
The low energy effective theory of Coulomb branch of a four dimensional $\mathcal{N}=2$ quantum field theory is described by a special K\"{a}hler geometry (SKG) \cite{Sierra:1983cc,gates1984superspace,de1984potentials,strominger1990special}. 
The determination of SKG is often solved by finding a Seiberg-Witten (SW) geometry, which 
involves a Seiberg-Witten curve fibering over Coulomb branch and an associated SW differential \cite{Seiberg:1994aj,Seiberg:1994rs}.

More generally, it is argued in \cite{Shapere:1999xr,Xie:2015rpa} that one can get SW geometry by using  three-fold fibration over the Coulomb branch. The construction goes as follows: let us start with an isolated rational three dimensional quasi-homogeneous hypersurface singularity $f:\C^4\rightarrow \C$.  This defines a four dimensional $\mathcal{N}=2$ superconformal field theory (SCFT). The SW geometry is expected \cite{Xie:2015rpa} to exist on the mini-versal deformation of the singularity $F(z,\lambda)$. One important missing ingredient in \cite{Xie:2015rpa} is that the SW differential was not specified. 

Over the mini-versal deformation space of the singularity, there exists special families of volume forms called \emph{primitive forms}. This is a deep notion introduced by K. Saito  \cite{saito1983period} by solving a version of Birkhoff factorization problem that arises from studying Hodge theory of isolated singularities. It was later realized that primitive forms play fundamental roles in 2d topological field theory \cite{Durbrovin96} : they produce Frobenius manifolds or solutions of WDVV equations for 2d Landau-Ginzburg models. See \cite{hertling2002frobenius} for an exposition on this topic. 

The purpose of this note 
is to show that the primitive form \cite{saito1983period} studied in 2d Landau-Ginzburg model context leads to the desired SW differential.  Our main statement about the SW differential of N=2 SCFT from three dimensional hypersurface singularity $f$ is  
$$
\boxed{\text{$\zeta$ primitive form}\Longrightarrow  \text{${\zeta\over dF}$ Seiberg-Witten differential}}. 
$$
Here $F$ is a miniversal deformation of $f$. $\zeta$ is a primitive form, which in our three-fold geometry is a special family of holomorphic $4$-form parametrized by the miniversal deformation. ${\zeta\over dF}$ is the \emph{Gelfand-Leray form} of $\zeta$, which gives a family of holomorphic $3$-form on each smooth Milnor fiber of $F$. In particular, the period map of ${\zeta\over dF}$ over vanishing homology leads to desired Seiberg-Witten geometry (see Section \ref{sec:SW}). The main support for the connection between primitive form and SW differential is about the integrability condition \eqref{theorem} for the existence of $\mathcal N=2$ prepotential arising from the Seiberg-Witten period map. The verification of such integrability requires understanding the relationship between the intersection pairing for vanishing homology and the period map. Primitive form provides precisely such a connection \cite{saito1983period} (see \eqref{eqn:intesection-form}).

The primitive form $\zeta$ is the analogue of the holomorphic $2$-form $\omega$ in the usual curve geometry of Seiberg-Witten. However, there is a difference between three-fold geometry and the curve geometry. In the curve case, the SW differential is given by a $1$-form $\lambda$ such that $\omega=d\lambda$. The analogue of $\lambda$ is the Gelfand-Leray form of $\zeta^{(-1)}$, which we call the first descendant  of $\zeta$ (see \eqref{eqn:first-descendant-GL}). The SW differential for three-fold geometry picks up $\zeta$ instead of $\zeta^{(-1)}$ for curve geometry, by the reason of shift of Hodge theory arising from the shift of dimension (see Section \ref{sec:SW} for details). In general, there exists infinite many descendant forms which is an important ingredient in K.Saito's theory of primitive form. They are captured by the Brieskorn lattice (see Section \ref{sec: Brieskorn}), which physics connection to gravitational descendants were studied in the 2d Landau-Ginzburg context  \cite{losev1993descendants,losev1998hodge} and in the topological string field theory context \cite{costello2012quantum}.

In the case of 3-fold ADE singularities, the primitive form is unique and does not depend on deformation parameters. In other words, it takes the form 
$$
\zeta= dx_1\wedge dx_2\wedge dx_3\wedge dx_4
$$
where $x_i$'s are linear holomorphic coordinates on $\C^4$. The associated SW differential takes the familiar form in the literature \cite{Shapere:1999xr}. The same statement holds when irrelevant deformations are not present (see Section \ref{sec: example}). This includes a large class of four dimensional $\mathcal{N}=2$ SCFT by compactifying 6d (2, 0) theory on a Riemann surface and then engineering using the 3-fold singularity. Beyond ADE singularities and with generic coupling turned on, the primitive form may not be unique and it varies nontrivially with respect to the deformation parameter, whose closed formulae are generally unknown (see Section \ref{sec: example} for some examples). This raises the interesting question on determining the precise SW differentials. Our result extends the Seiberg-Witten solution to include irrelevant deformations in this general context. 

This note is organized as follows: Section 2 reviews the construction of 4d $\mathcal{N}=2$ SCFT using singularity theory, and we discuss the integrability condition that 
a SW differential has to satisfy; Section 3 introduces the notion of Brieskorn lattice and primitive forms in singularity theory. We show that the Gelfand-Leray form of primitive forms solve the integrability condition of SW 
differentials and present several examples of primitive forms.

\section{Singularity and Rigid special K\"{a}hler structure}
In this section we review the construction of 4d $\mathcal{N}=2$ SCFT using singularity theory \cite{Shapere:1999xr,Xie:2015rpa}. We give a geometric formulation of special K\"{a}hler geometry that will be convenient to our study of primitive period map for Milnor fibrations.

\subsection{Singularity and 4d $\mathcal{N}=2$ SCFT}\label{sec: singularity and N=2 SCFT}
We start with an isolated three-fold hypersurface singularity $f:\C^4\rightarrow \C$ with an effective $\C^*$-action:
\begin{equation}
f(\lambda^{q_i} x_i)=\lambda f(x_i),~~q_i>0, \quad \lambda \in \C^*. 
\label{cstar}
\end{equation}
Such $f$ is called \emph{quasi-homogeneous}, and $q_i$ is the weight of $x_i$. The number
$$
\hat c_f=\sum_{i}(1-2q_i)
$$
is the central charge of the associated two dimensional $(2,2)$ SCFT defined by a Landau-Ginzburg model with superpotential $f$.  

To define a 4d $\mathcal{N}=2$ SCFT, we need 
 $$
 \sum q_i >1.\quad (\text{or equivalently}\  \hat c_f<2).
 $$ 
The SW geometry is described by the mini-versal deformation of the singularity:
\begin{equation}
F(x_i,\lambda)=f(x_i)+\sum_{\alpha=1}^\mu \lambda_\alpha \phi_\alpha: \C^4\times M\to \C.
\end{equation}
Here $\phi_\alpha$ is the basis of the Jacobi algebra $\Jac(f)$:
\begin{equation}
\Jac(f)={\C[x_1,\ldots, x_4]/ \bracket{ {\partial f\over \partial x_1}, \ldots, {\partial f\over \partial x_4}}}.
\end{equation}
$\mu=\dim_{\C}\Jac(f)$ is called the \emph{Milnor number}. The mini-versal deformation space $M$ is locally parametrized by $\lambda_\alpha$. One can define a scaling on base parameter $\lambda_\alpha$: 
\begin{equation}
[\lambda_\alpha]={1-Q_\alpha\over \sum q_i -1}.
\end{equation}
Here $Q_\alpha$ is the charge of $\phi_\alpha$ under $\C^*$ action defined in \eqref{cstar}.
Using above scalings, the base parameters are separated into three categories:
\begin{itemize}
\item $[\lambda_\alpha]<1$: Coupling constants 
\item $[\lambda_\alpha]=1$: Mass parameters whose number is denoted as $m_f$
\item $[\lambda_\alpha]>1$: Expectation value of Coulomb branch operators whose number is denoted as $r_f$.
\end{itemize}
Coupling constants and mass parameters are UV parameters while expectation value of Coulomb branch operators denote 
different vacua. If we order the polynomials $\phi_\alpha$ in terms of the quasi-homogeneous weight, then  we find a paring between the parameters: 
\begin{equation}
[\lambda_i]+[\lambda_{\mu-i}]=2.
\end{equation}
Because of this pairing, we have $\mu = 2r_f +m_f$. At a generic point parameterized by $\lambda$, the low energy theory is described by a $U(1)^{r_f}$ abelian gauge theory, and the effective theory is 
described by effective photon coupling $\tau_{ij}(\lambda)$ whose determination is a central task of understanding a four dimensional $\mathcal{N}=2$ theory.

Strictly speaking, the above discussion is valid at the UV point $\lambda=0$. When we deform away to arrive at a generic singularity $F$, we have to use a set of special \emph{flat coordinates} $\tau_\alpha$ on $M$ to parametrize the mini-versal deformation space. The flat coordinates $\tau_\alpha$ has the same rescaling weight of $\lambda_\alpha$ and they coincide at the first order 
\begin{equation}
   \lambda_\alpha=\tau_\alpha+O(\tau^2). 
\end{equation}
The above description of Coulomb parameters, mass parameters and coupling constants should use the flat coordinates $\tau_\alpha$ instead at a generic point of mini-versal deformation.  

 The existence of such parameters is a general consequence of K. Saito's theory of primitive form \cite{saito1983period}. For ADE singularities, the primitive form is the trivial volume form $\zeta= dx_1\wedge dx_2\wedge dx_3 \wedge dx_4$ \cite{saito1983period}, and the flat coordinates can be understood from the study of the oscillatory integral over Lefschetz thimbles (for example \eqref{eqn: QDE})
$$
   \int e^{F/t}\zeta. 
$$
In this case, flat coordinates are explicitly known \cite{dijkgraaf1991topological} \cite{Durbrovin96}. Beyond ADE and simple elliptic singularities, explicit formulae of primitive form and flat coordinates are generally unknown. Nevertheless, they can be computed perturbatively in order of $\lambda_\alpha$ \cite{li2013primitive, li2014mirror}. 

We define a Coulomb slice $U\subset M$ to be a subspace of $M$ where the coupling constants and mass parameters (in terms of flat coordinates $\tau_\alpha$) are fixed to be constants. $U$ is parametrized by the expectation value of Coulomb branch operators and has dimension $r_f$. Our goal is to establish Seiberg-Witten geometry on each Coulomb slice.

\subsection{Seiberg-Witten geometry}

The low energy effective theory defines a rigid special K\"{a}hler geometry, which can be defined in many ways. Here we use the following descriptions:

\textbf{Definition 1}: There are $2r$ holomorphic one forms \cite{Craps:1997gp}: $e_\alpha^A dz^{\alpha},~~A=1,\ldots,r,~~h_\alpha^A dz^{\alpha},~~A=1,\ldots,r,$ such that  $U_\alpha=(e_\alpha^A, h_\alpha^B)$ satisfies:
\begin{equation}
\bar{\partial}_{\bar{\alpha}} U_\beta =0,~~\partial_{[\alpha}U_{\beta]}=0.
\label{crucial1}
\end{equation}
and 
\begin{equation}
\langle U_\alpha, U_\beta\rangle =0,~~g_{\alpha \bar{\beta}}=i\langle U_\alpha, \bar{U}_{\bar{\beta}} \rangle.
\label{crucial2}
\end{equation}
Here the symplectic bracket is defined as 
\begin{equation}
\langle U_\alpha, U_\beta\rangle = e_\alpha^A h_\beta^A-h_\alpha^A e_\beta^A,~~\langle U_\alpha, \bar{U}_{\bar{\beta}} \rangle= e_\alpha^A \bar{h}_{\bar{\beta}}^{A}-h_\alpha^A \bar{e}_{\bar{\beta}}^A.
\end{equation}
In overlapping region, the transition function is:
\begin{equation}
U_{\alpha,(i)} dz_{(i)}^\alpha = e^{ic_{ij}}M_{ij}U_{\alpha, (j)} d z_{(j)}^{\alpha}.
\end{equation}
with $c_{ij} \in R$ and $M_{ij}\in Sp(2r, R)$.

\textbf{Definition 2}:
We have a $2r$ dimensional holomorphic symplectic manifold $X$ with a holomorphic symplectic $(2,0)$  form $\omega$, and a map $\pi:X\rightarrow B$ such that the fibre is Lagrangian, i.e.
the restriction of $\omega$ on fibre is zero. So the fibre is an abelian variety and we also impose a polarization on the fibre which is given by a real positive two form $\eta$. The K\"{a}hler metric 
on the base $B$ is given by integrating form $\eta^{r-1}\wedge \omega \wedge \bar{\omega}$ over the fibre \cite{Donagi:1995cf}. 

The special K\"{a}hler geometry can be often found by using a curve fibration over the Coulomb branch and an associated Seiberg-Witten differential. For a rank $g$ theory, its SW geometry 
might be specified by finding a genus $g$ Riemann surface firbration over a g dimensional subspace of its deformation space (which has dimension $3g-3,~g\geq 2$), and by finding a SW differential $\lambda$. 
Once the SW curve and SW differential $\lambda$ are found, one 
can find low energy effective theory by studying the period integral of $\lambda$ (definition 1 of SKG), and one can also translate 
the period integral interpretation in terms of an abelian variety fibration (definition 2 of SKG) . There are many examples of above type in the literature, however, there seems to be no systematical understanding.

In our case, we have a 3-fold fibration, and the SW differential is expected to be a three form. 
Let us now discuss what is the constraint from special K\"{a}hler geometry on volume form which we also called Seiberg-Witten differential. 
For simplicity, assume the vanishing of mass parameters so that the intersection pairing on the vanishing homology $H_3(F^{-1}(0),\Z)$ of the Milnor fiber is non-degenerate. Let us choose a symplectic basis $(A_i, B_j)$ such that their intersection numbers are given by 
$$
 A_i \cdot B_j=\delta_{ij}. 
$$
Our purpose is to find a 3-form $\Omega$ such that 
\begin{equation}
e_{\alpha}^i=\partial_{\tau_{\alpha}}\int_{A_i} \Omega,~~h_{\alpha}^j=\partial_{\tau_\alpha}\int_{B_j} \Omega
\end{equation}
gives rise to special K\"{a}hler geometry on each Coulomb slice (so the variation $\pa_{\tau_\alpha}$ is only along Coulomb moduli).  $\Omega$ will be called the SW differential in the context of singularity theory. The period integral is holomorphic and the equation \eqref{crucial1} is satisfied. The first equation implies that the following intersection pairing involving forms $\Omega$ vanishes:
\begin{equation}
\boxed{\langle \partial_{\tau_\alpha}\Omega, \partial_{\tau_\beta} \Omega \rangle =0}.
\label{theorem}
\end{equation}
Here $\abracket{-,-}$ is the induced intersection pairing on cohomology. $\Omega$ is viewed as giving a section of the vanishing cohomology and $\pa_{\tau_\alpha}\Omega$ is the derivative with respect to the Gauss-Manin connection along Coulomb moduli. This condition puts severe constrain on the choice of $\Omega$. 

Geometrically, let $\mathsf{H}$ denote the cohomology $H^3(F^{-1}(0),\C)$ which are locally identified  (away from the discriminant locus) for each Milnor fiber in terms of the flat Gauss-Manin connection. Assume for simplicity the vanishing of mass parameters, so $\mathsf{H}$ is a natural symplectic space whose symplectic pairing is induced by the intersection pairing. The period map of $\Omega$ over the vanishing cycles can be viewed as locally defining a map 
$$
  [\Omega]: M\to \mathsf{H}, \quad \tau_\alpha\to \int \Omega(\tau).
$$
A choice of sympletic basis allows us to decompose into isotropic subspaces 
$$
   \mathsf{H}=\mathsf{H}_+\oplus \mathsf{H}_-
$$
where $\mathsf{H}_+$ are dual to A-cycles and $\mathsf{H}_-$ are dual to B-cycles. In particular, we can identify 
$$
 \mathsf{H}=T^* \mathsf{H}_+
$$
as the cotangent bundle of $\mathsf{H}_+$. Let $a_i, a^D_i$ denote linear coordinates on $\mathsf{H}_+, \mathsf{H}_-$ that are dual to $A, B$-cycles. Following \cite{Seiberg:1994rs}, we denote the natural holomorphic 2-form 
$$
   \omega=  \sum_i da_i \wedge da^D_i
$$
and symplectic 2-form 
$$
 \eta= \Im  \sum_i \bracket{da^D_i \wedge d\bar a_i }. 
$$
$\omega$ and $\eta$ in fact does not depend on the choice of coordinates associated to A,B-cycles.

Let $U\subset M$ be a Coulomb slice. Let us denote the restriction of the above period map to Coulomb slice by
$$
\mathcal P: U\to \mathsf{H}. 
$$
Then condition \eqref{theorem} says that 
$$
  \mathcal P^* \omega=0. 
$$
Together with primitivity condition, it implies that $\mathcal P$ embeds $U$ locally into a Lagrangian submanifold of $\mathsf{H}$. Under the identification of $\mathsf{H}$ with $T^*\mathsf{H}_+$, we find holomorphic function $\mathcal F_0$ on $\mathsf{H}_+$ such that
$$
\mathcal P(U)=\text{Graph}(d\mathcal F_0)
$$
is given by the graph of the 1-form $d\mathcal F_0$ viewed as a section of $T^*\mathsf{H}_+$. Then $\mathcal F_0$ is the Seiberg-Witten prepotential on the Coulomb slice $U$. We expect that $\mathcal P^* \eta$ would give a K\"{a}hler form on $U$. It would be an interesting problem to justify this positivity condition. 

\section{Primitive form and Seiberg-Witten geometry}
In this section, we explain K. Saito's theory of primitive forms \cite{saito1983period} associated to isolated singularities and establish its relation with Seiberg-Witten geometry.  We show that the Seiberg-Witten differential is given by the Gelfand-Leray form of K. Saito's primitive form.

\subsection{Jacobian algebra and residue}
Let $f: (\C^{n+1}, \0)\to (\C,\0)$ be a holomorphic function with an isolated singularity at the origin $\0$. Let $X$ denote the germ of $\C^{n+1}$ at $\0$ with coordinates $\{x_i\}$ and $\Delta$ be the germ of $\C$ at $\0$.  Let $\Omega^k_X$ be holomorphic differential  $k$-forms on $X$. An element $\alpha$ of $\Omega^k_X$ is represented by 
$$
   \alpha=\sum_{i_1,\cdots,i_k} \alpha_{i_1,\cdots,i_k}(x)dx_{i_1}\wedge dx_{i_k}
$$
where $\alpha_{i_1,\cdots,i_k}(x)$'s are holomorphic functions around $\0$.  We consider the quotient space 
\begin{align}
 \boxed{\Omega_f:= \Omega^{n+1}_X/ df\wedge \Omega^{n}_X}.
\end{align}
Elements of $\Omega_f$ are given by equivalence class of $(n+1)$-forms on $X$. Given $\alpha\in \Omega^{n+1}_X$, we write $[\alpha]$ for its class in $\Omega_f$. Then for two elements $\alpha_1, \alpha_2\in \Omega^{n+1}_X$, they define the same class in $\Omega_f$ if and only if they differ by the form $df\wedge \beta$ for some $\beta\in \Omega^n_X$:
$$
  [\alpha_1]=[\alpha_2] \ \text{in\ }\Omega_f \quad \Longleftrightarrow \quad \alpha_1=\alpha_2+df \wedge \beta \quad \text{for some $\beta\in \Omega^n_X$}.
  \label{equiva}
 $$

Let $\C\{x_1,\cdots, x_n,x_{n+1}\}$ be the germ of holomorphic functions at $\0$. Let us define 
\begin{align}
   \Jac(f):= \C\{x_1,\cdots, x_{n+1}\}/ (\pa_if)
\end{align}
the quotient of $\C\{x_1,\cdots, x_n\}$ by relations generated by the partial derivatives $\pa_{1}f, \cdots, \pa_{n+1}f$. $\Jac(f)$ is called the Jacobian algebra of $f$. The assumption that $\0$ being an isolated singularity of $f$ implies the Milnor number
\begin{align}
\mu=\dim_{\C}\Jac(f)<\infty.
\end{align}
It is easy to see that we can identify 
\begin{align}
   \Omega_f= \Jac(f) d \xx,  \quad d \xx=dx_1\wedge \cdots\wedge dx_{n+1}. 
\end{align}
In fact,  a $n+1$ form is specified by a holomorphic function $g$, and a $n$ form is specified by a vector $h_i$. The equivalence relation in \eqref{equiva} implies that 
\begin{equation*}
g_1=g_2+\sum_{i=1}^{n+1}{\partial f\over \partial x_i} h_i,
\end{equation*}
which coincides with the  equivalence relation in Jacobian algebra.

In this paper, we will mainly focus on the case when $f$ is a polynomial. In this case we can replace $\C\{x_1,\cdots, x_{n+1}\}$ by the polynomial ring $\C[x_1,\cdots,x_{n+1}]$ and identify 
\begin{align}
 \Jac(f)=\C[x_1,\cdots,x_{n+1}]/(\pa_if).
\end{align}

Intrinsically, there exists a non-degenerate pairing given by the residue at $\0$ (see \cite{griffiths2014principles})
\begin{align}
  \Res_f: \Omega_f\otimes \Omega_f \to \C
\end{align}
where 
\begin{align}
     \Res_f(g d\xx, h d\xx)\propto{1\over (2\pi i)^{n+1}}\int_{\Gamma} \left( g h d\xx\over \pa_1f \cdots \pa_{n+1}f\right).
\end{align}
Here $\Gamma$ is a cycle in $X$ defined by $\{ |\pa_1 f(x)|=\cdots=|\pa_{n+1}f(x)|=\epsilon \}$ for $\epsilon>0$ a small enough positive number. We have left with a freedom of normalization constant. 

Let us consider the case when $f$ is a quasi-homogeneous polynomial 
\begin{align}
   f(\lambda^{q_i}x_i)=\lambda f(x_i), \quad \lambda \in \C^*. 
\end{align}
This $\C^*$-action induces a natural homogeneous degree on $\C[x_i]$ and $\Jac(f)$ denoted by $\deg$: 
\begin{align}
\deg(x_1^{k_1}\cdots x_{n+1}^{k_{n+1}})= k_1 q_1+\cdots+ k_{n+1}q_{n+1}. 
\end{align}
It turns out that we can choose a basis of $\Jac(f)$ represented by homogeneous polynomials $\{\phi_\alpha\}_{\alpha=1,
\cdots,\mu}$ such that
\begin{align*}
   0=\deg(\phi_1)< \deg(\phi_2)\leq \cdots \leq \deg(\phi_{\mu-1})< \deg(\phi_\mu)=\hat c_f, \quad
       \deg(\phi_i)+\deg(\phi_{\mu-i})=\hat c_f. 
\end{align*}
where 
\begin{align}
  \hat c_f=\sum_{i=1}^{n+1}(1-2q_i).
\end{align}
The residue pairing is completely determined by the ring structure of $\Jac(f)$ and the normalization constant on the top degree element $\phi_\mu$
$$
   \Res_f(d\xx, \phi_\mu d\xx)=C\neq 0.
$$
In fact, let $g, h\in \Jac(f)$ be two homogeneous elements. Then 
$$
\Res_f(g d\xx, h d\xx)=\begin{cases}
0 & \text{if}\  \deg(g)+\deg(h)\neq \hat c_f\\
m C & \text{if}\ gh =m \phi_\mu\ \text{in}\  \Jac(f). 
\end{cases}
$$
A useful consequence is that 
\begin{align}\label{eqn:vanishing-residue}
\Res_f(g d\xx, h d\xx)=0, \quad \text{if}\quad \deg (g)+\deg(h)<\hat c_f. 
\end{align}
This will be crucial for us to obtain Seiberg-Witten differential in singularity theory. 

\subsection{Brieskorn lattice and descendant forms}\label{sec: Brieskorn}
The space $\Omega_f$ is the leading term of the \textbf{Brieskorn lattice}
\begin{align}
\boxed{\mc{H}_f^{(0)}:= \Omega^{n+1}_X/ df\wedge d\Omega^{n-1}_X}
\end{align}
which plays an important role in the Hodge theory of singularities. Here we take a quotient of $\Omega^{n+1}_X$ by elements of the form $df\wedge d\beta$ for some $(n-1)$-form $\beta$. There is a natural surjection
$$ 
   \mc{H}_f^{(0)}\to \Omega_f. 
$$

 $\mc{H}_f^{(0)}$ carries important analytic structures that capture various properties of vanishing cycles associated to the Milnor fibration $f$. To see this, let us introduce another quotient 
 \begin{align}
\boxed{\mc{H}_f:= \Omega^{n+1}_X[z^{-1}]/ (d+z^{-1}df\wedge) \Omega^n_X[z^{-1}]}.
\end{align}
Here $z$ is new variable, and $\Omega^{k}[z^{-1}]$ are $k$-forms valued in the polynomial ring $\C[z^{-1}]$. An element $\alpha$ of $\Omega^{k}[z^{-1}]$ is represented by a finite sum 
$$
\alpha=\alpha_0 +\alpha_1 z^{-1}+\cdots+ \alpha_m z^{-m}, \quad \alpha_j\in \Omega^{k}, \quad m\geq 0. 
$$
Note that we do not include positive powers in $z$ here. The space $\mc{H}_f$ is called the \emph{Gauss-Manin system}.  Geometrically, an element $[\alpha]\in \mc H_f$ represents the de Rham class of the oscillatory integral (paired with Lefschetz thimbles)
$$
    \int e^{f/z}\alpha,
$$
and the quotient by $d+z^{-1}df$ represents integration by parts
$$
 \int e^{f/z} (d\beta+ z^{-1}df\wedge \beta)=\int d\bracket{e^{f/z}\beta}=0. 
$$
Consider the natural map 
$$
 \varphi:  \Omega^{n+1}_X\to \mc{H}_f
$$
by sending a $(n+1)$-form to its class in $\mc{H}_f$. We claim that 
\begin{align}
\boxed{\ker \varphi= df\wedge d\Omega^{n-1}_X}. 
\end{align}
In fact, $df\wedge d\Omega^{n-1}_X$ lies in $\ker \varphi$ since for any $\beta \in \Omega^{n-1}_X$, 
$$
   df\wedge d\beta =-(d+z^{-1}df\wedge)(df\wedge \beta) 
$$
hence zero in $\mc{H}_f$.

 On the other hand,  let $\alpha\in \ker \varphi$. Then there exists $n$-forms $\beta_i$ such that
 $$
   \alpha= (d+z^{-1}df\wedge)(\beta_0+z^{-1}\beta_1+\cdots z^{-k}\beta_k).
 $$
Comparing each order in $z$, this is equivalent to 
\begin{align*}
   \alpha&= d\beta_0\\
   d\beta_1&= -df \wedge \beta_0\\
    & \cdots\\
   d\beta_{k}&=-df\wedge \beta_{k-1}\\
      0 &= -df\wedge \beta_k. 
\end{align*}
The last equation implies that $\beta_k=df\wedge \gamma_k$ for some $(n-1)$-form $\gamma_k$ since the complex
$$
    0\rightarrow\Omega^{0}_X \stackrel{df\wedge}{\to}\Omega^{1}_X \stackrel{df\wedge}{\to}\cdots \Omega^{n+1}_X\rightarrow \Omega_f\rightarrow 0
$$
is exact around $\0$. Then $df\wedge (\beta_{k-1}+d\gamma_k)=0$, which implies the existence of $(n-1)$-form $\gamma_{k-1}$ such that $\beta_{k-1}=-d\gamma_k+df\wedge \gamma_{k-1}$. Working backwards we find $\alpha\in df\wedge d\Omega^{n-1}_X$.

As a consequence, $\varphi$ induces an embedding 
\begin{align}
    \boxed{\mc{H}_f^{(0)}\into \mc{H}_f}. 
\end{align}

$\mc{H}_f$ carries a natural action by multiplying $z^{-1}$. Less obviously, this operation is in fact invertible on $\mc{H}_f$! In other words, the inverse of $z^{-1}$-multiplication, which we simply call the $z$-multiplication, is also well-defined on $\mc{H}_f$, even though positive powers of $z$ have not appeared in our definition. The operator of $z$-multiplication is defined as follows. 

For $\alpha = z^{-1}\alpha_1+\cdots+ z^{-k}\alpha_k, k\geq 1$, we define
$$
z \cdot (\alpha):=z^{0}\alpha_1+\cdots+ z^{-k+1}\alpha_k.
$$
The nontrivial part is to see how the $z$-multiplication acts on $\mc{H}_f^{(0)}$. 

Let $\alpha \in \mc{H}_f^{(0)}$. Since $\alpha$ is a top form, there exists $\beta$ such that
$$
  \alpha =d\beta. 
$$
Then the $z$-multiplication on $\alpha$ is defined by 
\begin{align}
  z \cdot (\alpha):= -df\wedge \beta\in \mc{H}_f^{(0)}.
\end{align}
Its class in $\mc{H}_f^{(0)}$ doesn't depend on the choice of $\beta$. In fact, if we have $\alpha=d\beta^\prime$, then there exists $\gamma$ such that $\beta^\prime=\beta+d\gamma$. We have
$$
  df\wedge \beta^\prime=df\wedge \beta+df\wedge d\gamma
$$
which defines the same equivalence class in $\mc{H}_f^{(0)}$. 
 Symbolically, 
\begin{align}
\boxed{z=  -{df\over d}}: \alpha \to -df\wedge d^{-1}(\alpha). 
\end{align}

Therefore we have a well-defined operator 
\begin{align}
\boxed{z: \mc{H}_f^{(0)}\to \mc{H}_f^{(0)}, \quad \mc{H}_f \to \mc H_f. }
\end{align}
It is easy to check that such defined $z$-multiplication operator is inverse to the  manifest $z^{-1}$-multiplication on $\mc H_f$,  justifying the name. 

If follows from the above description that $\Omega_f$ is precisely the quotient
\begin{align}
  \boxed{\Omega_f= \mc{H}_f^{(0)}/ z\mc{H}_f^{(0)}}. 
\end{align}

Moreover, there exists a covariant derivative $\nabla_z$ defined on $\mc{H}_f$ by 
\begin{align}
  \nabla_z: z^{-k}\alpha\to -k z^{-k-1}\alpha-f z^{-k-2}\alpha, \quad k\geq 0. 
\end{align}
It  has a manifest meaning in terms of oscillatory integral 
\begin{align}
\pa_z \int e^{f/z}\alpha= \int e^{f/z}\nabla_z \alpha. 
\end{align}

\noindent \textbf{Remark}: The differential structure on $\mc{H}_f$ in $z$-variable is precisely the Laplace transform of the Gauss-Manin connection associated to the local system of vanishing cohomologies for the Milnor fibration $f: X\to \Delta$. For reader's convenience, we identify our notations 
\begin{align}\label{eqn:Laplace}
   \delta_1(=\pa_{t_1})=-z^{-1}, \quad t_1=-z^2 \nabla_z
\end{align}
where $t_1, \delta_1$ are the notations used in \cite{saito1983period}. Then $t_1$ can be viewed as the coordinate for the image $\Delta$ of $f: X\to \Delta$. Geometrically, the Laplace transform connects oscillatory integral with period integral by 
\begin{align}
  \int_{\Gamma} e^{f/z} \alpha= \int  dt_1  e^{t_1/z}\int_{\Gamma_{t_1}} {\alpha\over df}. 
\end{align}
Here $\Gamma_{t_1}$ is the intersection of $\Gamma$ with $f^{-1}(t_1)$. ${\alpha\over df}$ is the {Gelfand-Leray form} \cite{arnolʹd2012singularities} of $\alpha$.  As a nontrivial analytic result of this fact, for any power series $h(z)=\sum_{k\geq 0}a_k z^k$ in $z$ such that its Borel transformation $\sum_{k\geq 0}a_k {z^k\over k!}$ is convergent at $z=0$, the above $z$-multiplication extends to a well-defined $h(z)$-multiplication on $\mc H_f$. 

\noindent \textbf{Remark}: If we forget about the analytic structure, then the formal completion of $\mc{H}_f^{(0)}$ and $\mc{H}_f$ with respect to the $z$-adic topology are given by \cite{saito1983period}
$$
\widehat{\mc{H}}_f^{(0)}= \Omega^{n+1}_X[[z]]/ (zd+df\wedge) \Omega^n_X[[z]], \quad \widehat{\mc{H}}_f= \Omega^{n+1}_X((z))/ (zd+df\wedge) \Omega^n_X((z)).
$$
Here $\Omega^{n+1}_X[[z]]$ are $(n+1)$-forms valued in formal power series in $z$, and $\Omega^{n+1}_X((z))$ are $(n+1)$-forms valued in Laurent series in $z$. The $z$-multiplication becomes manifest in these expressions, and we still have 
$
  {\Omega_f= \widehat{\mc{H}}_f^{(0)}/ z\widehat{\mc{H}}_f^{(0)}}. 
$

$$
$$
\noindent \textbf{Definition}: For any integer $k$, we define the subspace of $\mc{H}_f$
\begin{align}
\boxed{\mc{H}_f^{(-k)}:= z^k \mc{H}_f^{(0)}}. 
\end{align}
Given $\omega\in \mc{H}_f^{(0)}$, we define its \textbf{$k$-th descendant} $\omega^{(-k)}\in \mc{H}_f$ by 
\begin{align}\label{defn-descendant}
\boxed{\omega^{(-k)}:= (-z)^k \omega}. 
\end{align}
In particular, a $(n+1)$-form $\omega$ determines an infinite tower of elements $\{\omega, \omega^{(-1)}, \omega^{(-2)}, \cdots\}$ in the Brieskorn lattice. Note that the $k$-th descendant has the meaning of $k$-th gravitational descendant in 2d Landau-Ginzburg models \cite{losev1993descendants,losev1998hodge,li2013primitive}.

Let us take a closer look at $\mc{H}_f^{(-1)}$. We have seen above that $\mc{H}_f^{(0)}$ is relevant for the oscillatory integral.  The space $\mc{H}_f^{(-1)}$ is in fact relevant for the period map over vanishing cycles. To see this, we observe that there is a natural isomorphism  \cite{saito1983period}
\begin{align}
  j:  \Omega^n_X/ (d \Omega^{n-1}_X+ df\wedge \Omega^{n-1}_X)  \iso \mc{H}_f^{(-1)}, \quad \beta \to df\wedge \beta. 
\end{align}

\noindent \textbf{Ramark}. In fact,  \cite{saito1983period} uses $\Omega^n_X/ (d \Omega^{n-1}_X+ df\wedge \Omega^{n-1}_X) $ to define $\mc{H}_f^{(-1)}$. In this paper, our convention is to view $\mc{H}_f^{(-1)}$ as a subspace of Brieskorn lattice for convenient. 

Let $\mathsf{H}_n$ be the flat bundle over $\Delta-\{\0\}$ associated to the local system of vanishing cycles. Its fiber over $p$ is given by 
$
H_n(f^{-1}(p);\C).
$
Let $\mathsf{H}^n$ be the dual cohomology bundle. For each element $\alpha \in \Omega^n_X/ (d \Omega^{n-1}_X+ df\wedge \Omega^{n-1}_X) $, it gives rise to a holomorphic section of $\mathsf{H}^n$ by restricting $\alpha$ to each Milnor fiber $f^{-1}(p)$. This defines a map 
$$
\Omega^n_X/ (d \Omega^{n-1}_X+ df\wedge \Omega^{n-1}_X) \to \Gamma(\Delta-\{\0\}, \mathsf{H}^n). 
$$

 Let $\omega\in \mc{H}_f^{(0)}$. We can write $\omega=d\Omega$ for some $n$-form $\Omega$. Since 
 \begin{align}\label{eqn:first-descendant-GL}
   \omega^{(-1)}= df \wedge \Omega, 
 \end{align}
 we see that $\omega^{(-1)}$ gives a well-defined section $[\Omega]$ of $\mathsf{H}^n$. This is essentially the period map. $\Omega$ is the {Gelfand-Leray form} of $\omega^{(-1)}$. It is worthwhile to note the analogue of the first descendant form with the algebraic integrable system for Seiberg-Witten geometry \cite{Donagi:1995cf}. 

On the other hand, the {Gelfand-Leray form} of $\omega$ also gives a section of $H^n$ 
$$
  t_1\to   \int  {\omega\over df}.
$$ 
The relationship of these two periods is 
\begin{align}
   \boxed{ \int_{\gamma_{t_1}}{\omega\over df}= \pa_{t_1} \int_{\gamma_{t_1}} \Omega}, \quad t_1\in \Delta-\{\0\}.
\end{align}
Here $\gamma_{t_1}$ is a flat family of cycle classes. This equation precisely reflects the identification of Laplace dual variables in \eqref{eqn:Laplace}.

\subsection{Primitive period map}
\subsubsection*{Primitive form}

Let $F: X\times M\to \Delta$ be a universal unfolding of $f$ parametrized by $M$. Let $o\in M$ be a reference point such that $f= F|_{X\times \{o\}}$. We will mainly consider the germ around $o$. Usually, $F$ can be presented by 
\begin{align}
  F(x,\lambda)= f(x)+ \sum_{\alpha=1}^\mu \lambda^\alpha \phi_\alpha(x)
\end{align}
where $\{\phi_\alpha\}$ is a basis of $\Jac(f)$ and $\{\lambda^\alpha\}$ gives local coordinates on $M$. 

We can similarly define $\Omega_F, \mc H_F, \mc H_F^{(0)}$, which are family versions of $f$ parametrized by $M$. For example, 
\begin{align}
  \mc H_F:= \Omega^{n+1}_{X\times M/M}[z^{-1}]/ (d+{z^{-1}dF})
\end{align}
where $\Omega^{n+1}_{X\times M/M}$ are relative $(n+1)$-forms with respect to the projection $X\times M\to M$. $d$ means the de Rham differential along $X$. $\mc H_F$ is relevant for the oscillatory integral 
$$
\int e^{F/z}(-). 
$$
The family version of the Brieskorn lattice is 
\begin{align}
\mc H_F^{(0)}= \Omega^{n+1}_{X\times M/M}/ dF\wedge d \Omega^{n-1}_{X\times M/M}
\end{align}
which is embedded into the Gauss-Manin system
$$
\mc H_F^{(0)}\into \mc H_F. 
$$
The manifest $z^{-1}$-multiplication is invertible on $\mc H_F$ and we can define the $z$-multiplication
\begin{align}
   z: \mc H_F^{(0)}\to \mc H_F^{(0)}, \quad \alpha\to -dF\wedge d^{-1}(\alpha). 
\end{align}
We still denote 
\begin{align}
   \mc H_F^{(-k)}= z^k \mc H_F^{(0)}, \quad \Omega_F= \mc H_F^{(0)}/z \mc H_F^{(0)}. 
\end{align}
Then the first descendant $\mc H_F^{(-1)}$ is relevant for the period integral over vanishing cycles. $\Omega_F$ forms a bundle of rank $\mu$ over $M$, equipped with a non-degenerate inner product given by the residue pairing
\begin{align}
\boxed{\eta_F: \Omega_F\times \Omega_F\to \underline{\C}_M}
\end{align}
where $\underline{\C}_M$ represents the trivial complex line bundle on $M$. 

There exists the flat Gauss-Manin connection on $ \mc H_F$ which allows us to take derivative along vector fields on $M$. Explicitly, 
\begin{align}
    \nabla_{V}  &\bbracket{g d\xx}:=\bbracket{ (\pa_V g+ {\pa_V F\over z})d\xx}, \quad V\in \Gamma(M, T_M).
\end{align}
It has a manifest meaning in terms of oscillatory integral 
$$
\pa_V \int e^{F/z} \alpha= \int e^{F/z} \nabla_V \alpha
$$
and satisfies the following Griffiths transversality condition 
$$
\nabla_V:  \mc H_F^{(-k)}\to  \mc H_F^{(-k+1)}, \quad V\in \Gamma(M, T_M).
$$

In \cite{saito1983period}, K. Saito defines a notion of primitive form as a special element of $\mc H_F^{(0)}$. Its precise mathematical definition is quite involved and come from a solution of Birkhoff factorization problem. We sketch some key properties here. Let $\zeta$ be a primitve form. 

First of all, if we choose a basis $v_1, \cdots, v_\mu$ of $T_M$, then the variations
$$
    \nabla_{v_1}\zeta^{(-1)}, \cdots, \nabla_{v_\mu} \zeta^{(-1)}
$$
lie in $\mc H_F^{(0)}$ and their projection to $\Omega_F$ form a local basis of $\Omega_F$. This allows us to define a non-degenerate pairing $g^\zeta$ on $T_M$
\begin{align}
     \boxed{g^\zeta(V_1, V_2):= \eta_F(\nabla_{V_1}\zeta^{(-1)}, \nabla_{V_2}\zeta^{(-1)})}, \quad V_i\in T_M
\end{align}
where  we have projected $\nabla_{V_i}\zeta^{(-1)}$ to $\Omega_F$ and take the residue pairing $\eta_F$.  The smart choice of primitive form $\zeta$ implies that the pairing $g^\zeta$ defines a torsion-free and flat holomorphic metric on $T_M$. In particular, there exists special flat coordinates $\{\tau^\alpha\}$ on $M$ such that 
\begin{align}
   \eta_{\alpha\beta}:=g^{\zeta}(\pa_{\tau^\alpha}, \pa_{\tau^\beta})
\end{align}
are constants.  The flat coordinates $\tau^\alpha$ are different from the unfolding parameters $\lambda^\alpha$ above and can be chosen to match at the first order
\begin{align}
 \boxed{\lambda^\alpha=\tau^\alpha +O(\tau^2)}. 
\end{align}
The deformation $F$ usually takes a complicated from in flat coordinates $\tau^\alpha$
$$
  {F(x, \tau)=f(x)+ \sum_{\alpha=1}^\mu \tau^\alpha \phi_\alpha(x) + O(\tau^2). }
$$

Secondly, in terms of flat coordinates $\{\tau^\alpha\}$, the geometric oscillatory integral of the primitive form satisfies the quantum differential equation 
\begin{align}\label{eqn: QDE}
 \boxed{\bracket{ \pa_{\tau^\alpha}\pa_{\tau^\beta}-{1\over z}A_{\alpha\beta}^\gamma(\tau) \pa_{\tau^\gamma} }\int e^{F/z}\zeta=0}. 
\end{align}
Here  $A_{\alpha\beta\gamma}(\tau)=A_{\alpha\beta}^\delta(\tau) \eta_{\delta \gamma}$ is the Yukawa coupling of 2d Landau-Ginzburg model that satisfies the WDVV equation.  

Thirdly, there exists an endomorphism $\NN\in \text{End}(T_M)$ which is flat with respect to the Levi-Civita connection associated to $g^\zeta$. Eigenvalues of $\NN$ are called the  ``exponents" of $f$. It is compatible with the flat inner product $g^{\zeta}$ in the sense that
$$
g^{\zeta}(\NN(v_1),v_2)+ g^{\zeta}(v_1, \NN(v_2))=(n+1)g^{\zeta}(v_1, v_2), \quad v_i\in T_M. 
$$
In particular, the linear map $\NN-{n+1\over 2}$ on $T_M$ is skew-symmetric with respect to $g^\zeta$. 

When $f$ is a quasi-homogeneous polynomial, the exponents lie between $[r, n+1-r]$ where $r=\sum_i q_i$. $\NN$ is diagonalized by the flat coordinates $\tau^\alpha$ such that 
$$
  \NN \pa_{\tau^\alpha}=(r+Q_\alpha)\pa_{\tau^\alpha}. 
$$
Here $Q_\alpha$ is the charge (homogeneous weight) of $\phi_\alpha$. 

These data lead to the structure of Frobenius manifolds associated to isolated singularities. The above properties are justified with the help of \emph{higher residue pairings} \cite{saito1983period}, which can be viewed as a special lifting of $\eta_F$ on $\Omega_F$ to the Brieskorn lattice $\mc H_F^{(0)}$. 

\subsubsection*{Period map}
Let us now consider the period map associated to a primitive form $\zeta$ \cite{saito1983period}. Let 
$$
 D\subset M
$$
be the discriminant locus of $F$. A point $\lambda \in D$ if the hypersurface $F(x;\lambda)=0$ in $X$ is singular. It is known that $D$ is an irreducible hypersurface.  Let
$$
  M^*= M-D
$$
be the complement of $D$. Over $M^*$, we have the flat homology bundle $\mathsf{H}_n$ of vanishing cycles and its dual cohomology bundle $\mathsf{H}^n$. We are interested in the period map of descendants of the primitive form $\zeta$
\begin{align}
 \bbracket{\zeta^{(-k)}\over dF}:  M^*\to \mathsf{H}^n.
\end{align}
Here $ \bbracket{\zeta^{(-k)}\over dF}$ means that we take the Gelfand-Leray form of $\zeta^{(-k)}$ to get a $n$-form on each hypersurface $F=0$ over $\lambda\in M^*$. Then its integration over vanishing cycles defines an element of the dual $\mathsf{H}^n$.

We specialize to the case of our interest when 
$$
n=2m+1
$$ 
is odd and $f$ quasi-homogeneous. We fix the flat coordinates $\tau^\alpha$ associated to $\zeta$. 

The first good property of primitive form is that the determinant \cite[(3.5.3)]{saito1983period}
$$
  \det \bracket{ {\pa\over \pa \tau^\alpha} \int_{\gamma_\beta(\lambda)} {\zeta^{({n-1\over 2}-1)}\over dF}}
$$
is constant and is nonzero if there exists no integer exponent in $(0,m]$. Here $\gamma_\beta(\lambda)$ is an integral basis of the vanishing homology. In particular, it is nonzero constant in our case of 3-fold hypersurface singularity $(m=1)$ when $\sum_{i=1}^4 {q_i}>1$. 

The second good property of primitive form is that the intersection form of the vanishing homology has a simple description in terms of primitive period map. In fact, the intersection pairing (which is skew-symmetric for $n$ odd)
$$
I=\abracket{-,-}: \mathsf{H}_n \times \mathsf{H}_n\to \C
$$
can be viewed as a flat skew-symmetric section $I$ of the bundle $\mathsf{H}^n\otimes \mathsf{H}^n$. In terms of the primitive form $\zeta$ and flat coordinates $\tau^\alpha$, it is given by \cite[(3.2.8)(3.4.1)]{saito1983period} (up to a constant)
\begin{align}
  I \propto \eta^{\alpha\beta} \bbracket{\nabla_{\pa_{\tau^\alpha}}\zeta^{(m-1)}\over dF} \otimes  \bbracket{  \nabla_{(\N-m-1)\pa_{\tau^\beta}}\zeta^{(m-1)}\over dF}, \quad m={n-1\over 2}.
\end{align} 
Here $(\N-m-1)\pa_{\tau^\beta})=(r+Q_\beta-m-1)\pa_{\tau^\beta}$ is the vector field obtained by applying the operator $(\N-m-1)$ to the vector field $\pa_{\tau^\beta}$. $\eta^{\alpha\beta}$ is the inverse matrix of $\eta_{\alpha\beta}$. 

In other words, for any two vanishing cycles $\gamma_1, \gamma_2$ over a Milnor fiber, we have 
\begin{align}\label{eqn:intesection-form}
\abracket{\gamma_1, \gamma_2}\propto \eta^{\alpha\beta} \bracket{\pa_{\tau^\alpha}\int_{\gamma_1} {\zeta^{(m-1)}\over dF} } \bracket{  ( (\N-m-1)\pa_{\tau^\beta})\int_{\gamma_2} {\zeta^{(m-1)}\over dF}}, \quad m={n-1\over 2}..
\end{align}
In the case when the intersection is non-degenerate, we have an induced intersection pairing on $\mathsf{H}^n$, still denoted by $\abracket{-,-}$. It follows from the above formula that 
\begin{align}\label{eqn:intersection-cohomology}
\boxed{
\abracket{ \bbracket{\nabla_{\pa_{\tau^\alpha}}\zeta^{(m-1)}\over dF} , \bbracket{\nabla_{\pa_{\tau^\beta}}\zeta^{(m-1)}\over dF} }=\eta_{F}\bracket{ \nabla_{\pa_{\tau^\alpha}} \zeta^{(-1)}, \nabla_{(\N-m-1)^{-1}\pa_{\tau^\beta}} \zeta^{(-1)}}, \quad m={n-1\over 2}.
}
\end{align}
Here $(\N-m-1)^{-1}$ is the inverse of $(\N-m-1)$, and we have used the fact that
$$
\eta_{F}\bracket{ \nabla_{\pa_{\tau^\alpha}} \zeta^{(-1)}, \nabla_{\pa_{\tau^\beta}} \zeta^{(-1)}}=\eta_{\alpha\beta}.
$$

\subsection{Seiberg-Witten differential} \label{sec:SW}
We are ready to discuss the Seiberg-Witten geometry of 3-fold hypersurface singularity. We work with quasi-homogeneous polynomial $f(x_1,x_2,x_3,x_4)$ 
$$
  n=3, \quad r=\sum q_i>1.
$$
Eigenvalues of $\NN$ take values in $[r, 4-r]$. Let us assume the vanishing of mass deformations. This implies the non-degeneracy of intersection pairing on vanishing homology.  We are interested in finding the appropriate period map such that \eqref{theorem} holds on Coulomb slice. 

Let $\tau^\alpha=\lambda^\alpha+O(\lambda^2)$ be flat coordinates on $M$ associated to a primitive form $\zeta$. In terms of 4d rescaling weight, we denote 
$$
  [\tau^\alpha]= {1-Q_\alpha\over r-1}. 
$$
\begin{itemize}
\item Parameters with $[\tau^\alpha]>1$ are Coulomb moduli. This is equivalent to  $Q_\alpha<{\hat c_f\over 2}$. 
\item Parameters with $[\tau^\alpha]<1$ are coupling constants. This is equivalent to  $Q_\alpha>{\hat c_f\over 2}$. 
\item Parameters with $[\tau^\alpha]=1$ are mass parameters. This is equivalent to  $Q_\alpha={\hat c_f\over 2}$. 
\end{itemize}

\noindent \textbf{Remark}: Note that we have ``corrected" the  notions from Section \ref{sec: singularity and N=2 SCFT} using flat coordinates $\tau^\alpha$ instead $\lambda^\alpha$. 

Let $U\subset M^*$ is a Coulomb slice of $M$. On $U$,  the coupling constants are fixed to be constants and $U$ is parametrized by the Coulomb moduli (we have assumed the vanishing of mass parameters here).  Let us consider the period map 
\begin{align}
 \mathcal P:   U \to \bbracket{\zeta\over dF} \in \mathsf{H}^n. 
\end{align}
 Let us consider the intersection pairing of the tangent map. By \eqref{eqn:intersection-cohomology} ($m=1$), we find
$$
  \abracket{\pa_{\tau^\alpha}\mathcal P, \pa_{\tau^\beta} \mathcal P}\propto \eta_F  \left( \pa_{\tau^\alpha} \zeta^{(-1)}, \left(({ \NN-2})^{-1}\pa_{\tau^\beta} \right)\zeta^{(-1)} \right)= {\eta_{\alpha\beta} \over r+Q_\beta-2}
$$
which is in fact a constant. At the undeformed polynomial $f$ where $\tau=0$, we have 
$$
 \eta_{\alpha\beta}= \Res_f(\phi_\alpha d\xx, \phi_\beta d \xx)
$$
where $\phi_\alpha, \phi_\beta$ corresponds to the deformation of the Coulomb parameter $\tau^\alpha, \tau^\beta$. Since 
$$
\deg(\phi_\alpha)+\deg(\phi_\beta)=Q_\alpha+Q_\beta<\hat c_f
$$
by the description of Coulomb moduli, we have by \eqref{eqn:vanishing-residue}
$$
\Res_f(\phi_\alpha d\xx, \phi_\beta d \xx)=0.
$$
As a consequence, the period map $\mathcal P$ satisfies the integrability condition \eqref{theorem} and embeds $U$ locally as a Lagrangian submanifold of $\mathsf{H}^n$. This strongly supports that
\begin{align}
\boxed{\text{$\zeta$ primitive form}\Longrightarrow  \text{$\bbracket{\zeta\over dF}$ Seiberg-Witten differential}}. 
\end{align}

\noindent \textbf{Remark}: We remark on the difference between three-fold singularity ($n=3$) and the Seiberg-Witten curve case ($n=1$). When $n=1$, the intersection formula \eqref{eqn:intersection-cohomology} becomes 
\begin{align}
{
\abracket{ \bbracket{\nabla_{\pa_{\tau^\alpha}}\zeta^{(-1)}\over dF} , \bbracket{\nabla_{\pa_{\tau^\beta}}\zeta^{(-1)}\over dF} }=\eta_{F}\bracket{ \nabla_{\pa_{\tau^\alpha}} \zeta^{(-1)}, \nabla_{(\N-1)^{-1}\pa_{\tau^\beta}} \zeta^{(-1)}}.
}
\end{align}
This suggests that the appropriate choice of Seiberg-Witten differential 
$$
     \bbracket{\zeta^{(-1)}\over dF}= \bbracket{d^{-1}\zeta}
$$
is given by a form $\Omega$ such that $d\Omega=\zeta$. This is indeed the basic structure of algebraic integrable systems associated to Seiberg-Witten curves. 

When we lift curve to three-fold by adding two more variables, the identification of higher residue pairings and periods requires a shift of the descendant  \cite[(2.4.2)]{saito1983period}.  This is the reason we find $\zeta=\zeta^{(0)}$ instead of $\zeta^{(-1)}$ in the 3-fold case. 

\subsection{Examples}\label{sec: example}
\subsubsection*{ADE singularity}  We consider 
$$
f(x)= x_1^2+ x_2^2+ x_3^ k+ x_4^N 
$$
with 
$$
  {1\over k}+{1\over N}>{1\over 2}. 
$$
This condition is equivalent to 
$$
  \hat c_f<1
$$
which corresponds to ADE singularities. This is the case when marginal deformations (zero scaling parameters) and irrelevant deformations (negative scaling parameters) do not appear.  A basis of $\Jac(f)$ is given by 
$$
  \{x_3^i  x_4^j\}_{0\leq i\leq k-2, 0\leq j\leq N-2}.
$$ 
It leads to an unfolding $F$ as described above. 
For ADE singularities, the primitive form is unique and is given by the trivial volume form \cite{saito1983period}
$$
\zeta= dx_1\wedge dx_2\wedge dx_3\wedge dx_4.
$$
Note that $\zeta$ does not depend on the deformation parameters, which is a special property of ADE singularities. 

\subsubsection*{Simple elliptic singularity}
We consider 
$$
f= x_1^3+ x_2^3+ x_3^3+x_4^2. 
$$
It has $\hat c_f=1$ and is one of the simple elliptic singularities. This is the case when irrelevant deformations (negative scaling parameters) do not appear. 

We consider the miniversal deformation 
$$
  F=f+ \lambda_1 + \lambda_2 x_1+\lambda_3 x_2+\lambda_4 x_3+\lambda_5 x_1x_2+\lambda_6 x_2x_3+\lambda_7 x_3x_1+\lambda_8 x_1x_2 x_3. 
$$
Primitive form of this example is nontrivial and is not unique. They depend only on the marginal parameter $\lambda_8$ described as follows. Consider the elliptic curve in $\mathbb P^2$ defined by 
$$
 E_{\lambda_8}=\{x_1^3+ x_2^3+ x_3^3+ \lambda_8 x_1x_2x_3=0\}\subset \mathbb P^2. 
$$
Let $\Omega_{\lambda_8}$ be the holomorphic volume form on $E_{\lambda_8}$ obtained by the residue 
$$
\Omega_{\lambda_8}=\Res_{E_{\lambda_8}} \bracket{x_1 dx_2dx_3- x_2 dx_1dx_3+x_3dx_1 dx_2\over x_1^3+ x_2^3+ x_3^3+ \lambda_8 x_1x_2x_3}.
$$
Here the form inside the bracket of the above equation is viewed as a rational 2-form on $\mathbb P^2$ in terms of homogenous coordinate with an order one pole along $E_{\lambda_8}$.  Its residue gives a holomorphic 1-form $\Omega_{\lambda_8}$ on  $E_{\lambda_8}$.

Let $\gamma$ be a flat family of $1$-cycles in $H_1(E_{\lambda_8}, \Z)$. It determines a primitive form by \cite{saito1983period}
$$
\zeta= {dx_1\wedge dx_2\wedge dx_3\wedge dx_4\over \int_{\gamma} \Omega_{\lambda_8}}. 
$$
In particular, the primitive form only depends on the parameter $\lambda_8$. Since $H_1(E_{\lambda_8}, \Z)$ has rank $2$, we see that the moduli space of primitive forms (up to rescaling) is one-dimensional. The corresponding SW differential is given by 
$$
\Omega={1\over \int_{\gamma} \Omega_{\lambda_8}} {dx_1\wedge dx_2\wedge dx_3\wedge dx_4 \over  dF}.
$$
Note that $\lambda_8$ is a coupling constant which is fixed on each Coulomb slice. Therefore the Seiberg-Witten geometry on the Coulomb slice can be safely taken to be the naive one
$$
  \Omega\to  {dx_1 \wedge dx_2 \wedge dx_3\wedge dx_4\over dF}.
$$

\subsubsection*{$E_{12}$-singularity}
We consider 
$$
f=x_1^2+x_2^2+x_3^3+x_4^7.
$$
This is type $E_{12}$ of the unimodular exceptional singularities, which is also called the $(A_2, A_6)$ theory in physics. The miniversal deformation is the following
\begin{align}
& F(x,\lambda)=x_1^2+x_2^2+x_3^3+x_4^7+\lambda_1+\lambda_2 x_4 +\lambda_3 x_4^2+\lambda_4 x_3 +\lambda_5 x_4^3+\lambda_6 x_3 x_4+\lambda_7 x_4^4+\lambda_8 x_3 x_4^2 \nonumber\\
& +\lambda_{9} x_4^5+\lambda_{10} x_3x_4^3+\lambda_{11} x_3x_4^4+\lambda_{12} x_3x_4^5.
\end{align}

This example contains an irrelevant deformation parameter $\lambda_{12}$, leading to a nontrivial story. There exists a unique primitive form for this example, but its closed formula is unknown. Nevertherless, there exists a perturbative algorithm \cite{li2013primitive, li2014mirror} to  compute the primitive form and flat coordinates order by order in deformation pamameters.  Up to order $10$, it is computed in \cite{li2013primitive} by
$$
\zeta=  (\varphi(x,\lambda)+O(\lambda^{11}) )dx_1\wedge dx_2\wedge dx_3\wedge dx_4.
$$
where 
\begin{align*}
 \varphi(x,\lambda)&=1+{4\over 3\cdot 7^2}\lambda_{11}\lambda_{12}^2-{64\over 3\cdot 7^4}\lambda_{11}^2\lambda_{12}^4-{76\over 3^2\cdot 7^4}\lambda_{10}\lambda_{12}^5+{937\over  3^3\cdot 7^5}\lambda_9\lambda_{12}^6+{218072\over 3^4\cdot 5\cdot 7^6}\lambda_{11}^3\lambda_{12}^6\\
         &\qquad+{1272169\over 3^4\cdot 5\cdot 7^7}\lambda_{10}\lambda_{11}\lambda_{12}^7+{28751\over 3^4\cdot  7^7}\lambda_{8}\lambda_{12}^8-{1212158\over 3^4 \cdot 7^8}\lambda_9\lambda_{11}\lambda_{12}^8-{38380\over 3^3\cdot 7^8}\lambda_{7}\lambda_{12}^9\\
         &\quad+\big({1\over 7^2} \lambda_{12}^3-{ 101\over 5\cdot 7^4} \lambda_{11}\lambda_{12}^5+ {1588303\over 3^4\cdot 5\cdot 7^7}\lambda_{11}^2\lambda_{12}^7+
                {378083\over 3^4\cdot 5\cdot 7^7}\lambda_{10}\lambda_{12}^8-{ 108144\over 3\cdot 7^8}\lambda_{9}\lambda_{12}^9\big)x_3\\
         &\quad+\big({1447\over 3^3\cdot 7^6} \lambda_{12}^7-{ 71290\over 3^3\cdot 7^8}\lambda_{11}\lambda_{12}^9\big)x_4   -{45434\over 3^4\cdot 7^8} \lambda_{12}^{10}x_3x_4\\
         &\quad -\big({53\over 3^2\cdot 7^4}\lambda_{12}^6-{ 46244\over 3^3\cdot 7^7}\lambda_{11}\lambda_{12}^8\big) x_3^2 +{22054\over 3^4\cdot 7^7}\lambda_{12}^9 x_3^3.
\end{align*}
In particular, the Seiberg-Witten differential of this example is not given by a rescaling of the trivial form and depends on the coupling constants in a nontrivial way. 

\subsubsection*{Vanishing irrelevant deformation}
We consider the Seiberg-Witten differential on the Coulomb slice $U_0\subset M^*$ where irrelevant deformations ($[\tau^\alpha]<0$ or $Q_\alpha> 1$) are zero. The restriction of primitive form on this slice is greatly simplified. On $U_0$, we only turn on deformations $\phi_\alpha$ whose charges $Q_\alpha$ satisfy
$$
   Q_\alpha\leq 1. \quad (\text{for Coulomb moduli}\ Q_\alpha< {\hat c_f}/2).
$$
This is similar to the ADE  and simple elliptic cases. An analogue of degree argument  (see for example \cite{saito1983period,li2013primitive}) implies that 
$$
\zeta|_{U_0}= {dx_1 \wedge dx_2\wedge dx_3 \wedge dx_4 \over P},
$$
where $P$ is a function of the marginal deformation parameters ($Q_\alpha=1$). In particular, the Seiberg-Witten differential on each slice can be taken to be the familiar form 
$$
{dx_1 \wedge dx_2\wedge dx_3 \wedge dx_4\over dF}.
$$

\subsection{Seiberg-Witten curve geometry revisited}

Let us now consider curve singularity, and the Seiberg-Witten differential is given by the Gelfand-Leray  form of the first decedent of primitive form (see Remark in Section \ref{sec:SW}). Concretely, we are looking for a one form $\lambda$ such that $d\lambda = \zeta$ with $\zeta$
the primitive form. If the primitive form is trivial, i.e. $\zeta= d p \wedge dx$, then $\lambda =p dx$ takes the familiar form in the literature. This is indeed the case when we turn off irrelevant deformations as discussed above. In general, the primitive form is nontrivial. 

Let us now compare our result with the known results in the literature. One can construct a large class of four dimensional 
$\mathcal{N}=2$ SCFT by compactifying 6d $(2,0)$ theory on a Riemann surface with regular and irregular singularity \cite{Gaiotto:2009we, Xie:2012hs}. The SW curve is identified with 
the spectral curve $det (p-\Phi(x))=0$, here $p$ is the coordinate on the cotangent bundle of the Riemann surface and $x$ is the coordinate of Riemann surface, 
and the SW differential is always the naive one $\lambda = p dx$. This seems in contradiction to our proposal that the primitive form is nontrivial and therefore the SW differential is complicated.
The resolution for the puzzle is following: the deformation with negative scaling dimension is suppressed in the Hitchin system description, therefore 
irrelevant deformations are not present, and indeed the SW differential is the naive one! Our result suggests that the SW geometry is much more complicated if 
we turn on irrelevant deformations.

\section{Conclusion}
We have shown that the primitive forms lead to SW differentials for theory defined by three-fold isolated hypersurface singularity with a $\C^*$ action. This construction 
provides another way of producing Seiberg-Witten geometry besides the usual Seiberg-Witten curve construction. Let us make some further remarks here:
\begin{itemize}
\item Our study suggests that one should identify the UV deformation parameters of the field theory with the special coordinates $\tau^\alpha$, and not 
the natural coordinate $\lambda^\alpha$ in the description of mini-versal deformation. 

\item Primitive form is in general different from the the naive volume form used in the literature. The complication comes from the irrelevant deformations corresponding to generators of 
Coulomb branch operators whose scaling dimension is bigger than two, and it is interesting to further understand the physical meaning of these irrelevant deformations \footnote{D.Xie would like to thank P.Argyres for the discussion on this point. } 
Moreover, it exists for any isolated singularity \cite{saito1989structure} and for a large class of Laurent polynomials with non-isolated singularities \cite{douai2003gauss,douai2004gauss}. 

\item The primitive form is not unique in general. Simple elliptic singularity is such an example. They correspond to the choice of monodromy compatible splittings of the Hodge filtration on Brieskorn lattices \cite{saito1983period,saito1989structure}. A natural one may be chosen either from the limiting mixed Hodge structure \cite{saito1989structure} or from the $tt^*$-geometry \cite{cecotti1991topological}. 

\item In this paper we have not explored the positivity of the induced metric from special K\"{a}hler geometry. This is a physically important property that is likely related to the  $tt^*$-geometry \cite{cecotti1991topological}. We hope to explore this in the future.  
\end{itemize}

We focus on $\mathcal{N}=2$  theory defined by 3d isolated hypersurface singularity with a $\C^*$ action in this paper. There are several possible interesting generalizations:
\begin{enumerate}
\item One can consider any quasi-homogeneous hypersurface singularity with central charge $\hat{c}<2$ and similarly study the role of primitive form. An example is a five dimensional hypersurface singularity $x_1^3+x_2^3+x_3^3+x_4^k+x_4x_5^2+x_6^2=0$ \cite{DelZotto:2015rca}. The analogue of Remark in Section \ref{sec:SW} suggests that the SW differential of this example is  $\zeta^{(1)}$. 
\item We can also consider 3d isolated rational hypersurface singularity without $\C^*$ action, and we should get a 4d $\mathcal{N}=2$ quantum field theory without conformal invariance. Primitive form exists in this case \cite{saito1989structure} and is still expected to give the SW differential.
\item It is interesting to study SW differential for other three dimensional rational Gorenstein singualrity, such as 3d rational quasi-homogeneous complete intersection singularity \cite{Chen:2016bzh,Wang:2016yha}.
\item One can also consider singularity defined using ${\C}^*$ variables (the SW solution of Yang-Mills theory involves such description). Primitive form exists for a large class \cite{douai2003gauss,douai2004gauss} of this case and it is interesting to study the corresponding SW differential. 
\end{enumerate}

\section*{Acknowledgements}
The work of S.T Yau is supported by  NSF grant  DMS-1159412, NSF grant PHY-
0937443, and NSF grant DMS-0804454.  
The work of DX is supported by Center for Mathematical Sciences and Applications at Harvard University, and in part by the Fundamental Laws Initiative of
the Center for the Fundamental Laws of Nature, Harvard University. SL is partially supported by Grant 20151080445 of Independent
Research Program at Tsinghua University. Part of this work was done when SL was visiting Center for Mathematical Sciences and Applications at Harvard University and Max Planck Institute for Mathematics in Jan 2018. SL thanks for their hospitality and provision of excellent working enviroment.


\bibliographystyle{utphys}

\bibliography{PLforRS}

\end{document}